\newcommand{\version}{February 6, 2018}
\documentclass[letterpaper,oneside,11pt,pdftex]{article}
\pdfoutput=1
\usepackage[bindingoffset=0.1cm,textheight=22.6cm,hdivide={*,16cm,*}, vdivide={*,22.5cm,*}]{geometry}
\usepackage{amsmath,amsfonts,amssymb}
\usepackage{mathtools}
\usepackage[T1]{fontenc}
\usepackage[utf8]{inputenc}
\usepackage{lmodern}
\usepackage{tensind}
\tensordelimiter{?}
\usepackage[numbers,square,comma,sort&compress]{natbib}
\usepackage[pdftex,hyperref,svgnames]{xcolor}
\usepackage[pdftex,bookmarksnumbered=true,breaklinks=true,colorlinks=true,linktocpage=true,linkcolor=MediumBlue,citecolor=ForestGreen,urlcolor=DarkRed]{hyperref}

\makeatletter
\AtBeginDocument{
  \hypersetup{
    pdfsubject = {\Abstract},
    pdfkeywords = {\keywords},
    pdftitle = {\@title},
    pdfauthor = {\@author}
  }
}
\makeatother

\makeatletter

 \@addtoreset{equation}{section}
 \makeatother

\renewcommand{\a}{\alpha}
\renewcommand{\b}{\beta}
\newcommand{\g}{\gamma}
\renewcommand{\d}{\delta}
\newcommand{\e}{\epsilon}

\renewcommand{\th}{\theta}

\newcommand{\m}{\mu}
\newcommand{\n}{\nu}

\newcommand{\s}{\sigma}
\renewcommand{\t}{\tau}

\newcommand{\ph}{\phi}

\newcommand{\w}{\omega}

\newcommand{\G}{\Gamma}

  \newcommand{\cO}{\mathcal{O}}

\newcommand{\pa}{\partial}

\newcommand{\sgn}[1]{\textrm{sgn}\!\left(#1\right)}
\newcommand{\nn}{\nonumber}
\newcommand{\eqnref}[1]{Eqn. \eqref{#1}}

\newcommand*{\vv}[1]{\vec{\mkern0mu#1}}
\newcommand{\ba}{{{\bar\alpha}}}
\newcommand{\txt}[1]{\textrm{#1}}
\newcommand{\tpfe}{thermoelastic-plastic flow equations}

\DeclarePairedDelimiter\abs{\lvert}{\rvert}
\newcommand{\coleq}{\vcentcolon=}

\title{\texorpdfstring{\begin{flushright}
        {\small LA-UR-16-24560}
       \end{flushright}\vspace{2em}}{}%
       Thermoelastic-Plastic Flow Equations in General Coordinates}

\author{Daniel N. Blaschke and Dean L. Preston}

\date{\version}

\newcommand{\Abstract}{%
The equations governing the thermoelastic-plastic flow of isotropic solids in the Prandtl-Reuss and small anisotropy approximations in Cartesian coordinates are generalized to arbitrary coordinate systems.
In applications the choice of coordinates is dictated by the symmetry of the solid flow.
The generally invariant equations are evaluated in spherical, cylindrical (including uniaxial), and both prolate and oblate spheroidal coordinates.
}

\newcommand{\keywords}{thermoelastic-plastic flow, metal plasticity}

\begin{document}

\maketitle

\thispagestyle{empty}
\begin{center}
\vspace{-0.3cm}
Los Alamos National Laboratory,
\\Los Alamos, NM, 87545, USA
\\[0.5cm]
\ttfamily{E-mail: dblaschke@lanl.gov, dean@lanl.gov}
\end{center}

\vspace{1.5em}

\begin{abstract}
\Abstract
\end{abstract}

\newpage
\tableofcontents


\section{Introduction}
\label{sec:intro}


The plastic deformation of a single crystal of a metal or alloy is mediated by the motion (glide) of dislocations~\cite{Hirth:1982}, which are linear topological crystal defects, in preferred glide planes and directions characteristic of the crystal structure.
Dislocation motion is accompanied by the relative slip of the crystal on opposite sides of the glide plane; the glide of a high density of dislocations under an applied stress is manifested at the macroscale as continuous plastic deformation.
The pairs of glide planes and directions along which the dislocation glide occurs are the so-called slip systems, which are specified by their Miller indices~\cite{Barrett:1966}.
Yielding on a given slip system is described by the Schmid Law~\cite{Schmid:1925}:
a single crystal yields in a given slip system $s$ only if the resolved (component of) shear stress achieves a critical value $\tau^s$, the yield strength of slip system $s$.
For each active slip system there is an associated hyperplane in stress space on which the resolved shear stress equals  $\tau^s$.
Single crystals yield on the stress space surface which is the inner envelop of all such hyperplanes, that is, on a polyhedron in stress space.
The theory and modeling of single crystal plasticity is an ongoing enterprise, see~\cite{Hansen:2013} and references therein.

In this paper the focus is on metal polycrystals, i.e. aggregates of single crystals or grains.
The grains in a typical metal polycrystal range in size from a few microns to about 1 millimeter.
On a length scale larger than the grain size the polycrystal is isotropic if the grains are randomly oriented.
However, plastic deformation of the polycrystal, for example, forming processes, can give the grains a preferred orientation, in which case the polycrystal is anisotropic.

In the continuum, or macroscopic, theory of polycrystal plasticity the microstructural features of the metal, such as grain size and the spatial distribution of dislocations, are represented by continuous variables or parameters that describe the internal state of the material \cite{Swearengen:1985a, Swearengen:1985b}.
The evolution of the internal state variables $\chi_i$, which can be scalars, vectors, or tensors, may be described by equations of the form
\begin{equation}
\dot{\chi}_i = g_i \left(\dot{\epsilon}^p_{jk}, \chi_j, T\right) \, ,
\end{equation}
where $\dot{\epsilon}^p_{jk}$ is the symmetric plastic strain rate tensor.
The strain rate  $\dot{\epsilon}^p_{jk}$ must be a function of the stress tensor $\tau_{ij}$, the temperature, $T$, and the internal state variables; $\tau_{ij}$ is the Cauchy stress, thus the $i$-component of the force on a surface element of area $dS$ with unit normal $\vec{n}$ is $\tau_{ij} \, n_j \, dS$.
The components of the stress tensor can be replaced by the equivalent set $\bar{P}$, $s_{ij}$ defined by
\begin{align}
\bar{P} &= - \frac{1}{3} \, \tau_{ii} \, , \nonumber\\
s_{ij} &= \tau_{ij} + \bar{P} \, \delta_{ij} \, ;
\label{eq:split-tau}
\end{align}
the $s_{ij}$ are the stress deviators and $\bar{P}$ is the mean compressive stress.
The stress deviation tensor is traceless, $s_{ii} = 0$, and $\bar{P}$ can be identified with the pressure only if all $s_{ij}$ vanish.
The condition for plastic yield may be written
\begin{equation}
f \left(s_{ij}, \dot{\epsilon}^p_{ij}, \chi_i, T, \bar{P} \right) = 0\, .
\label{eq:yieldsurf}
\end{equation}
The surface in deviatoric stress space defined by Equation \eqref{eq:yieldsurf} is the yield surface.
The well-known normality flow rule states that an incremental plastic strain is normal to the yield surface in $\tau_{ij}$ space.
It follows that
\begin{equation}
d \epsilon^p_{ij} = d \xi \, \frac{\partial f}{\partial \tau_{ij}} =  d \xi \, \left( \delta_{ki} \, \delta_{lj} -\frac{1}{3} \, \delta_{ij} \, \delta_{kl} \right) \,
\frac{\partial f}{\partial s_{kl}} \, ,
\label{eq:normalityflow}
\end{equation}
where $\xi$ is dimensionless.
In the case of an isotropic polycrystal, the yield function can depend on its vector and tensor arguments only through their rotational invariants.
In particular, it can depend on the stress deviation tensor only through its two rotational invariants $S_2 = (1/2) \, s_{ij} \, s_{ji}$ and $S_3 = (1/3) \, s_{ij} \, s_{jk} \, s_{ki}$.
An $S_3$-dependent yield function can result in various second-order phenomena \cite{Freudenthal:1969,Ronay:1967};
we assume that $f$ is independent of $S_3$.
Similarly, we assume that $f$ depends on the plastic strain rate only through the rotational invariant
\begin{equation}
\dot{\psi} = \left( \frac{2}{3} \, \dot{\epsilon}_{ij}^p \, \dot{\epsilon}_{ji}^p \right)^{1/2} \, ;
\label{eq:psidot}
\end{equation}
$\psi$ is the equivalent plastic strain.
The notation is that of Wallace \cite{Wallace:1985};
the equivalent plastic strain is denoted $\epsilon$ by many authors.
The simplest possible yield condition is then
\begin{equation}
f\left(S_2, \dot{\psi},\chi_k, T, \bar{P} \right)= \sqrt{S_2} - K\left(\dot{\psi}, \chi_i, T, \bar{P} \right) / \sqrt{2} = 0 \, .
\label{eq:vonMises1}
\end{equation}
This is the classical von Mises yield condition~\cite{Mises:1913} which in principal axes (those that diagonalize $\tau_{ij}$) reads
\begin{equation}
s_{11}^2+ s_{22}^2 + s_{33}^2 = K^2 \, ,
\label{eq:vonMises2}
\end{equation}
which is the equation of a sphere in deviatoric stress space.
Using Equation \eqref{eq:normalityflow} we obtain
\begin{equation}
d \epsilon^p_{ij} =  \frac{3}{4} \, \frac{s_{ij}}{\tau} \, d \psi \, ,
\label{eq:PR-approx}
\end{equation}
where
\begin{equation}
\tau^2 = \frac{3}{4} \, S_2\, .
\label{eq:tausq}
\end{equation}
This proportionality of the plastic strain increments to the stress deviators is known as the Prandtl-Reuss approximation~\cite{Prandtl:1925,Reuss:1930}.

Work hardening, i.e. the increase in yield strength with deformation, is represented by an increase in the radius $K$ of the von Mises yield sphere.
It can be modeled most simply by replacing the dependency of $K$ on the $\chi_i$ by a dependence on only $\psi$.
In effect, $\psi$ is used as an approximate internal state variable.
Replacing $\chi_i$ in \eqref{eq:vonMises1} with $\psi$, using \eqref{eq:tausq}, and then solving for $\dot{\psi}$ would give a constitutive relation of the form
\begin{equation}
\dot{\psi} = \dot{\psi} \, \left(\tau, \psi, T, \bar{P} \right) \, .
\label{eq:pcr}
\end{equation}
Since the microstructure of a metal depends on the deformation path and not just on the integrated plastic strain along that path, $\psi$ is certainly not a true state variable.
Nevertheless, a constitutive relation of this form for isotropic metals and alloys works very well in practice.
We shall use constitutive relation \eqref{eq:pcr} in this paper.

In accordance with \eqnref{eq:pcr} the plastic flow stress, defined as $\sigma = 2 \, \tau$, is measured in the laboratory as a function of $\psi$ for various choices of $T$ and $\dot{\psi}$.
These stress-strain curves can be measured by mechanical testing machines ($\dot{\psi} \leq 10^2 s^{-1}$), Hopkinson bars ($10^2 s^{-1} \leq \dot{\psi} \leq 10^4 s^{-1}$), and other techniques at higher strain rates.
For a review of experimental methods to determine the plastic flow properties of metals, see e.g.~\cite{Davison:1979,Field:2004} and references therein.

The description of a macroscopic solid flow process requires, in addition to the plastic constitutive relation, partial differential equations for mass, momentum, and energy conservation, entropy production, and $\bar{P}({\bf x}, t)$,  $s_{ij}({\bf x}, t)$, and $T({\bf x}, t)$.
The complete set of equations describing the plastic flow of an isotropic polycrystal in the Prandtl-Reuss approximation has been derived by Wallace in Cartesian coordinates \cite{Wallace:1970,Wallace:1985}.

Whenever a plastic flow process exhibits a certain symmetry it is most conveniently formulated in a coordinate system that respects that symmetry.
In this paper we generalize the thermoelastic-plastic flow equations from Cartesian to general coordinates.
Although some of the equations (e.g. momentum conservation) have been expressed in a more general form (see e.g.~\cite{Chitanvis:1997,Lubliner:2008}), the full set of thermoelastic-plastic flow equations  in general coordinates has never been published, but only appeared in an internal report by one of us \cite{Preston:1987}.
Given the generally covariant (invariant) equations, the flow equations in particular coordinate systems can be straightforwardly derived.
Here we evaluate the generally invariant equations in spherical, cylindrical (including uniaxial), and spheroidal coordinates;
these coordinates were chosen in view of their potential applications to modeling or experiments respecting those symmetries.
In particular, spheroidal coordinates could be used for the description of non-spherical void growth in ductile materials \cite{Gurson:1977,Gologanu:1993,Monchiet:2007,Keralavarma:2017};
previous theories of void growth have not taken into account the full set of thermoelastic-plastic flow equations.

\section{Derivation of thermoelastic-plastic flow equations}

We begin with a brief derivation of the thermoelastic-plastic flow equations of an isotropic material in the Prandtl-Reuss approximation following references \cite{Preston:1987,Wallace:1985} (see also~\cite{Wallace:1970,Wallace:1972}).

On length scales much greater than that of the microstructure, a polycrystal can be modeled as a continuum, hence it can be divided into infinitesimal mass elements.
The location of a mass element at time $t$ in some spatially fixed, i.e Eulerian, coordinate system is $x^i(t)$.
Alternatively, one may also choose to work with Lagrangian coordinates $X^i$ which are carried by the mass elements through the course of their motion and which serve as ``labels'' for the mass elements.
At some initial time $t_0$ the mass element may be located at $X^i=x^i(t_0)$.
Any material variable $Q$ may be expressed in terms of Eulerian or Lagrangian coordinates, i.e. $Q(x^i,t)$ or $Q(X^i,t)$ respectively.
The Eulerian and Lagrangian time derivatives are related via
\begin{align}
 \pa_t Q\big|_X&=\pa_t Q\big|_x+v^i\pa_i Q|_t \,, &
 v^i&=\pa_t x^i
 \,, \label{eq:timeder-EulLag}
\end{align}
where $\pa_t\coleq\frac{\pa}{\pa t}$ and $\pa_i\coleq\frac{\pa}{\pa x^i}$ (the latter always referring to Eulerian coordinates).
Introducing the transformation matrix between Eulerian and Lagrangian coordinates as
\begin{align}
 ?\a^i_j?\coleq\frac{\pa x^i}{\pa X^j}
 \,,
\end{align}
the relation between the Lagrangian and Eulerian spatial derivatives of a scalar quantity $Q$ follows from the chain rule
\footnote{For tensorial variables, the partial derivatives on both sides must be replaced by covariant derivatives.}:
\begin{align}
 \frac{\pa Q}{\pa X^i}=?\a^j_i?\pa_j Q
 \,. \label{eq:grad-EulLag}
\end{align}

\subsection{Mass and momentum conservation}

In order to derive the mass and momentum conservation equations for infinitesimal mass elements, we follow here a different approach than that employed in the material science literature:
We use relativistic notation to combine energy, momentum, and the stress tensor into the energy-momentum tensor $T^{\m\n}$  where $\m,\n\in 0,1,2,3$ and $0$ denotes the temporal coordinate $x^0=ct$ with $c$ the speed of light.
In general, it can be derived by varying the action (i.e. the space-time integral of the Lagrangian) with respect to the space-time metric (which may be set to the flat one after varying).
It follows that $T^{\m\n}$ is symmetric in its indices.
Furthermore, as a consequence of diffeomorphism invariance of the action it is covariantly conserved
\begin{align}
\nabla_\m T^{\m\n}=0\,, \qquad \n = 0, 1, 2, 3,
\label{eq:4dim-conservation}
\end{align}
as can be checked by applying Noether's theorem.
Here $\nabla_\m$ is the covariant derivative.
The covariant derivatives of a vector $v^{\a}$ and a (contravariant) second-rank tensor $ T^{\a \b}$ are
\begin{align}
 \nabla_\m v^{\a }&=\pa_{\m} v^{\a} + \G^{\a}_{\b \m} v^{\b}  \,, \nn\\
 \nabla_\m T^{\a \b} &=\pa_{\m} T^{\a \b} + \G^{\a}_{\g \m}  T^{\g \b} + \G^{\b}_{\g \m}  T^{\a \g} \,,
\end{align}
where $\G^{\a}_{\b \g}=\tfrac12g^{\a \m}\left(\pa_{\g} g_{\b \m}+\pa_{\b} g_{\g \m}-\pa_{\m} g_{\b \g}\right)$ are the usual Christoffel symbols with respect to the metric $g_{\a \b}$.
For a recent review and guide to the literature on the energy-momentum tensor and Noether's theorem see~\cite{Blaschke:2016ohs}.
Here we are, of course, interested in flat space-time in the non-relativistic limit where all velocities are much smaller than $c$.
However, we do want to derive our equations for arbitrary coordinates and therefore keep the general relativistic notation; see~\cite{Ohanian:2013} for an introduction to general relativity and the formalism we use in this section.
Thus, we may write for our infinitesimal mass element of density $\rho$ and speed $v^i=\pa_t x^i\! \ll c\,$:
\begin{align}
 T^{00}&=E=\rho c^2\,, &
 T^{0i}&=T^{i0}=\rho v^i c\,, &
 T^{ij}&=\rho v^iv^j-\t^{ij}
 \,,
\end{align}
where $i,j\in1,2,3$ and $\t^{ij}=\t^{ji}$ is the Cauchy stress.
\footnote{Symmetry of $\t^{ij}$ in its indices may also be checked explicitly by using angular momentum conservation~\cite{Wallace:1985}.}

Equation \eqref{eq:4dim-conservation} for $\n=0$ is the mass conservation equation (or energy conservation since $E=\rho c^2$) in general Eulerian coordinates.
Since $\pa_0=(1/c)\pa_t$ the speed of light cancels out and we find
\begin{align}
 \nabla_t\rho+\nabla_i(\rho v^i)=0 \, . 
 \label{eq:mass-cons-Eul}
\end{align}
Since the Christoffel symbol $\G^0_{00}$ vanishes we have $\nabla_t\rho=\pa_t\rho$, and using the relation \eqref{eq:timeder-EulLag} we may convert it to mixed Lagrangian/Eulerian form where the time derivative is Lagrangian:
\begin{align}
 \pa_t\rho\big|_X +\rho \nabla_i(v^i)=0
 \,. \label{eq:mass-cons-Lag}
\end{align}

Likewise, we derive the momentum conservation equation by considering $\n=i\in1,2,3$ which yields
\begin{align}
 \nabla_t(\rho v^j)+\nabla_i(\rho v^iv^j)-\nabla_i\t^{ij}=0
 \, \label{eq:momentum-cons0}
\end{align}
in Eulerian form.
Again, we have $\nabla_t\to\pa_t$ since the corresponding Christoffel symbols vanish (assuming the metric $g_{\m\n}$ satisfies $\pa_0 g_{\m\n}=0=g_{0i}$ and $g_{00}$ constant).
Let us examine the first term of the l.h.s. of \eqnref{eq:momentum-cons0} more closely:
\begin{align}
 \pa_t(\rho v^j)&=v^j\pa_t\rho +\rho \pa_t v^j=-v^j\nabla_i(\rho v^i)+\rho \pa_tv^j\nn\\
 &=\rho \left(\pa_tv^j + v^i \nabla_i v^j\right) -\nabla_i(\rho v^iv^j)
 \,,
\end{align}
where in the second step we used the mass conservation equation in Eulerian form, \eqnref{eq:mass-cons-Eul}.
Inserting this result into \eqnref{eq:momentum-cons0} finally yields
\begin{align}
 \rho\frac{Dv^j}{Dt}=\nabla_i\t^{ij}
 \,,
\end{align}
where $D / Dt$ is the absolute derivative of Brillouin~\cite{Brillouin:1964}
\begin{align}
 \frac{Dv^j}{Dt} &=\pa_tv^j\big|_x+v^i\nabla_i v^j \nn\\
&=\pa_tv^j\big|_X+\G^j_{lk}v^lv^k
\label{eq:def-total-time}
\end{align}
in Eulerian and Lagrangian form, respectively; we will generally use the latter.

\subsection{Displacements and strain}

The displacement of a mass element from its initial position is $u^i=x^i(X^j,t)-X^i$.
Deformation changes the initial separation of two material points from $dX^i$ to
\begin{align}
 dx^i&=dX^i+du^i=(\d^i_j+\bar\nabla_j u^i)dX^j 
 \,,
\end{align}
where $\bar\nabla_j$ denotes the covariant derivative with respect to the Lagrangian coordinate $X^j$.
We may therefore write the displacement gradients as
\begin{align}
 ?u^i_j?&\coleq\bar\nabla_j u^i
 =?{{\bar\a}}^i_j?-\d^i_j 
 \,,
\end{align}
which defines the tensor $?\ba^i_j?\coleq \bar\nabla_j x^i$.
Note that $?\ba^i_j?=?\a^i_j?$ in Cartesian coordinates, but not in general.
Hence
\begin{align}
 \abs{d\vv{x}}^2-\abs{d\vv{X}}^2=2\eta_{ij}dX^idX^j
 \,,
\end{align}
where
\begin{align}
 \eta_{ij}&=\frac12\left(g_{lk}?\ba^l_i??\ba^k_j?-g_{ij}\right)=\frac12\left(g_{il}?u^l_j?+g_{jl}?u^l_i?+g_{lk}?u^l_i??u^k_j?\right)
 \label{eq:murnaghantensor}
\end{align}
is the finite strain tensor of Murnaghan~\cite{Murnaghan:1937} (also known as Lagrangian strain or as the Green-Saint-Venant strain tensor~\cite{Lubliner:2008}) and $g_{ij}$ is the metric tensor for the coordinate system under consideration.
Note that \eqref{eq:murnaghantensor}, which is often used in the Lagrangian description of plasticity, is not the only definition of finite strain:
Another example (which we will not use here) is the Eulerian Almansi finite strain, which constitutes the Eulerian pendent to our present Lagrangian strain.
While both strain tensors mentioned above measure the change in length compared to the original length (which is referred to as ``engineering'' strain), it is sometimes useful to consider logarithmic (or ``natural'') strain instead~\cite{Lubliner:2008}.
For each of these strain measures there is a corresponding conjugate stress.
For example, the stress which is the conjugate to Lagrangian strain is the second Piola-Kirchhoff stress~\cite{Lubliner:2008} which can be easily transformed to the more commonly used Cauchy stress.

A further infinitesimal deformation $\d x^i=x'^i-x^i$ changes the separation $dx^i$ to
\begin{align}
 dx'^i&=\big(\d^i_j+\nabla_{\!j} \d x^i\big)dx^j \nn\\
 &=\big(\d^i_j+?u^i_j?+(\nabla_{\!k} \d x^i)(\d^k_j+?u^k_j?)\big)dX^j
 \,,
\end{align}
where $\nabla_{\!j} \d x^i=\pa_j \d x^i+\G^i_{jk}\d x^k$ denotes the covariant derivative with respect to Eulerian coordinates $x^j$ with Christoffel symbol $\G^i_{jk}$.
Hence to linear order in $\d x^i$ we have
\begin{align}
 \abs{d\vv{x}'}^2-\abs{d\vv{X}}^2&=2(\eta_{ij}+d\eta_{ij})dX^idX^j
 \,,
\end{align}
with
\begin{align}
 d\eta_{ij}&=\frac12{?\ba^m_i??\ba^n_j?}\big(g_{ml}\nabla_{\!n}\d x^l+g_{nk}\nabla_{\!m}\d x^k\big)
 \,. \label{eq:d-eta}
\end{align}
Furthermore, let us define the ``infinitesimal strain'' $\e_{ij}$ as the symmetric part of $?u^i_j?$
\begin{align}
 \e_{ij}&=\frac{1}2\left(g_{ik}?u^k_j?+g_{jk}?u^k_i?\right)
 \,. \label{eq:pure-strain}
\end{align}
Clearly, to linear order in the infinitesimal strain $?\e^i_j?$ we have $\eta_{ij}\approx \e_{ij}+\cO((\e_{ij})^2)$; see \eqref{eq:murnaghantensor}.
Although we will occasionally use the approximation $\eta_{ij}\approx \e_{ij}$ in our derivations below, the thermoelastic-plastic flow equations, which we summarize in \eqref{eq:geninv}, are generally valid for \emph{finite} strain~\cite{Wallace:1970}.

Finally, any second rank tensor $A^{ij}$ transforms under active infinitesimal rotations according to $dA^{ij}_\w=-g^{jl}A^{ik}d\w_{kl}+g^{il}d\w_{l_k}A^{kj}$, where
\begin{align}
 d\w_{ij}&=\frac12\left(g_{ik}\nabla_{\!j}\d x^k-g_{jk}\nabla_{\!i}\d x^k\right)
 \,. \label{eq:pure-rot}
\end{align}

\subsection{Energy conservation}

According to the first law of thermodynamics, a differential change in internal energy $U$ is the sum of the differential changes in work $dW$ (which may be divided into elastic and plastic contributions) and heat increment $dQ$:
\begin{align}
 dU=dW^\txt{e}+dW^\txt{p}+dQ
 \,. \label{eq:dU-initial}
\end{align}
Experiments teach us that only a small amount of plastic work goes into changing the defect structure of the material, i.e. most of the work is dissipated.
In order to simplify our equations we make the approximation that all plastic work is converted into heat
\footnote{Typically more than 87\% of the plastic work (depending on the material it may be much more) is dissipated~\cite{Farren:1925,Titchener:1958}.
However, the error we incur in the present approximation is worth the significant simplification of the theory~\cite{Wallace:1980}.}
and thus the total entropy increment is
\begin{align}
 TdS&=dW^\txt{p}+dQ
 \,. \label{eq:dS-initial}
\end{align}
Now consider a small region $R$ of material bounded by surface $S$.
An infinitesimal deformation of the material changes the coordinates of a point from $x^i$ to $x^i+\d x^i$.
The work done to effect this infinitesimal deformation is
\begin{align}
 dw&=\int_S \t^{ik}dS_k\d x^jg_{ij}=\int_R \nabla_k\big(\t^{ik}\d x^jg_{ij}\big)dV
 \,,
\end{align}
where we have absorbed the square root of the determinant of the metric inside $dV$.
Being metric compatible, the covariant derivative has the property $\nabla_k g_{ij}=0$.
If we now assume that $\nabla_i \t^{ij}\approx0$ for $S$ sufficiently small, which in Cartesian coordinates translates to the stress tensor being approximately constant over the small region $R$, we may write $dw=g_{ij}\t^{ik}\int_R \nabla_k\d x^jdV$.
Note that only the symmetric part of $\nabla_k\d x^j$ contributes because
the Cauchy stress
$\t^{ij}$ is a symmetric tensor.
Finally, if we assume that $\nabla_k\d x^j$ is nearly constant over $R$, we may divide the equation by the infinitesimal mass of $r$ and write the work per unit mass as
\footnote{
Note that the second Piola-Kirchhoff stress tensor $T^{mn}$ is related to the Cauchy stress $\t^{ij}$ via $T^{mn}=(\det\!\bar\a) ?{(\ba^{-1})}^m_i? ?{(\ba^{-1})}^n_j? \t^{ij}$ in our current notation, see e.g. \cite{Lubliner:2008} for definitions of the various stress tensors.
}
\begin{align}
 dW&=\frac{1}{2\rho}\t^{ij}\big(g_{jk}\nabla_i\d x^k+g_{ik}\nabla_j\d x^k\big) \nn\\
 &=\frac1\rho \t^{ij}?{(\ba^{-1})}^m_i??{(\ba^{-1})}^n_j?d\eta_{mn}
 \,, \label{eq:dW-initial}
\end{align}
where in the last step we have inserted \eqnref{eq:d-eta}.

Combining equations \eqref{eq:dU-initial}, \eqref{eq:dS-initial} and \eqref{eq:dW-initial} we may write the differential change in internal energy per unit mass as
\begin{align}
 \rho dU&=\t^{ij}?{(\ba^{-1}_\txt{e})}^m_i??{(\ba^{-1}_\txt{e})}^n_j?d\eta_{mn}^\txt{e}+\rho TdS
 \,, \label{eq:dU-semifinal}
\end{align}
where sub-/superscripts ``e'' denote elastic contributions.

It can be shown~\cite{Preston:1987} that all quantities are in fact independent of the chosen initial time $t_0$ (as they must be).
We therefore have the freedom of choosing our current configuration as our initial configuration, which means that $?\a^i_j?=\d^i_j=?\ba^i_j?$ (since $u^i\to0$) and $d\eta_{ij}\to d\e_{ij}$ where
\begin{align}
 d\e_{ij}&=\frac12\left(g_{ik}\nabla_{\!j}\d x^k+g_{jk}\nabla_{\!i}\d x^k\right)
 \,. \label{eq:depsilonij}
\end{align}
It follows from equations \eqref{eq:dU-semifinal} and \eqref{eq:depsilonij} that
\begin{align}
 \rho dU&=\t^{ij}d\e_{ij}^\txt{e}+\rho TdS
 \,. \label{eq:energy-cons}
\end{align}

\subsection{Stress tensor and temperature increments}

A differential change in $\t^{ij}$ is given by
\begin{align}
 d\t^{ij}&=\frac{\pa \t^{ij}}{\pa \eta^\txt{e}_{kl}}d\eta^\txt{e}_{kl}+\frac{\pa \t^{ij}}{\pa S}dS+d\t^{ij}_{\w}
 \,, \label{eq:stress-changes}
\end{align}
where
\begin{align}
 d\t^{ij}_{\w}=-\big(g^{jl}\t^{ik}d\w_{kl}+g^{il}\t^{jk}d\w_{l_k}\big)
 \,,
\end{align}
denotes the change in stress resulting from an infinitesimal rotation of the material; see \eqref{eq:pure-rot}.

The first term of \eqref{eq:stress-changes} can be computed from \eqref{eq:dU-semifinal} via
\begin{align}
 \t^{ij}&=\rho?{\ba_\txt{e}\!}^i_m??{\ba_\txt{e}\!}^j_n?\frac{\pa U(X^i,\eta^\txt{e}_{ij},S)}{\pa \eta^\txt{e}_{mn}}
 \,. \label{eq:tau-from-U}
\end{align}
In order to do so, we will need the relations
\begin{align}
 \frac{\pa ?\ba^i_j?}{\pa\eta_{kl}}&=g^{im}?{(\ba^{-1})}^{(k}_m?\d^{l)}_j
 =\frac{g^{im}}2\left[?{(\ba^{-1})}^k_m?\d^l_j+?{(\ba^{-1})}^l_m?\d^k_j\right] \,, \nn\\
 \frac{\pa\rho}{\pa\eta^\txt{e}_{ij}}&=-\rho?{(\ba^{-1}_\txt{e})}^i_k??{(\ba^{-1}_\txt{e})}^j_l?g^{kl}
 \,, \label{eq:relations}
\end{align}
where the second one follows from the mass conservation equation \eqref{eq:mass-cons-Lag} converted to pure Lagrangian form
\footnote{The easiest way to see this is by performing this short calculation in Cartesian coordinates.
The final result, written in the differential form above, then holds in any coordinates.},
and then written in differential form as $d\rho=-\rho?{(\ba^{-1})}^j_i?d?\ba^i_j?$.
In \eqnref{eq:relations} we introduced a shorthand notation for ``symmetrization in the indices'' using round brackets: $a_{(ij)}\coleq \frac12(a_{ij}+a_{ji})$.
We thus find
\begin{align}
 \frac{\pa\t^{ij}}{\pa \eta^\txt{e}_{kl}}&=\rho?{\ba_\txt{e}\!}^i_m??{\ba_\txt{e}\!}^j_n?\frac{\pa^2 U(X^i,\eta^\txt{e}_{ij},S)}{\pa \eta^\txt{e}_{kl}\pa \eta^\txt{e}_{mn}}+\left[g^{in}?{(\ba^{-1}_\txt{e})}^{(k}_n??{(\ba^{-1}_\txt{e})}^{l)}_m?\t^{mj}+(i\leftrightarrow j)\right] \nn\\
 &\quad -g^{mn}?{(\ba^{-1}_\txt{e})}^k_m??{(\ba^{-1}_\txt{e})}^l_n?\t^{ij}
 \,.
\end{align}
Choosing again our current configuration as our initial configuration, we obtain the adiabatic stress strain coefficients $B_s^{ijkl}$ to lowest order in the
infinitesimal strains
\begin{align}
 B_s^{ijkl}\coleq\frac{\pa\t^{ij}}{\pa \e^\txt{e}_{kl}}&=\rho\frac{\pa^2 U}{\pa \e^\txt{e}_{kl}\pa \e^\txt{e}_{ij}}+g^{i(k}\t^{l)j}+g^{j(k}\t^{l)i} 
 -g^{kl}\t^{ij}
 \,.
\end{align}
As their name indicates, these coefficients relate changes in strain to changes in stress at constant entropy.

The second term of \eqref{eq:stress-changes} can similarly be computed from \eqref{eq:tau-from-U} with the result
\begin{align}
 \frac{\pa \t^{ij}}{\pa S}&=-\rho\g^{ij}T\,, &
 \g^{ij}&=-\frac1T\frac{\pa^2 U}{\pa S\pa \e^\txt{e}_{ij}}
 \,, \label{eq:defgrueneisen}
\end{align}
where the coefficients $\g^{ij}$ are the anisotropic Gr{\"u}neisen parameters.
At constant elastic configuration ($d \epsilon^e_{ij}  = 0$) $dU = T dS$, hence the first equation in \eqnref{eq:defgrueneisen} may be written
\begin{align}
 \rho\g^{ij} &= - \frac{\pa \t^{ij}}{\pa U}\Big|_{\epsilon^\txt{e}}
 \,, \label{eq:gruneisen}
\end{align}
which shows that the Gr{\"u}neisen parameters quantify the changes in the stresses due to heating of the material.
Substituting the above results into \eqnref{eq:stress-changes} we get
\begin{align}
 d\t^{ij}&=B_s^{ijkl}d\e^\txt{e}_{kl}-\rho\g^{ij}TdS-\big(g^{jl}\t^{ik}d\w_{kl}+g^{il}\t^{jk}d\w_{l_k}\big)
 \,. \label{eq:stress-increments}
\end{align}


Finally, the change in temperature can be written as
\begin{align}
 dT&=\frac{\pa T}{\pa\eta^\txt{e}_{ij}}d\eta^\txt{e}_{ij}+\frac{\pa T}{\pa S}dS
 \,.
\end{align}
Using $T=\pa U / \pa S \big|_{\epsilon^\txt{e}} $,
which follows from \eqref{eq:dU-semifinal}, and the definition of the heat capacity at constant elastic configuration $C_{\epsilon^\txt{e}}$,
\begin{align}
 \frac{\pa T}{\pa S}\Big|_{\epsilon^\txt{e}}&=\frac{T}{C_{\epsilon^\txt{e}}}
 \,,
\end{align}
we find with the same assumptions as before and using the result \eqref{eq:defgrueneisen}
\begin{align}
 dT&=-T\g^{ij}d\e^\txt{e}_{ij}+\frac{T}{C_{\epsilon^\txt{e}}}dS
 \,. \label{eq:temp-increments}
\end{align}

\subsection{The small anisotropy approximation}

We now assume that the unstressed metal is isotropic.
However, in an applied anisotropic stress field the elastic strains are anisotropic hence the material is anisotropic.During elastic-plastic deformation the stress deviators must lie within or on the yield surface.
The magnitudes of the $s^{ij}$ are limited to values much less than the shear elastic constants.
Therefore the elastic strains are small and consequently the material anisotropy is small.
It then follows that the thermoelastic coefficients may be replaced by their values at zero elastic strain, i.e.  $\g^{ij}\to\g g^{ij}$, $C_{\e^\txt{e}}\to C_V$ (the constant volume heat capacity), and the $B_s^{ijkl}$ become linear combinations of $B$ and $G$, the bulk and shear moduli respectively (see references~\cite{Wallace:1970,Wallace:1980,Wallace:1985} for details).
These replacements will be referred to as the small anisotropy approximation

\subsection{Generally invariant thermoelastic-plastic flow equations }

We now derive the generally invariant (form invariance under arbitrary coordinate transformations) thermoelastic-plastic flow equations assuming small material anisotropy and an $S_3$-independent yield function, i.e. the Prandtl-Reuss approximation.
They are derived from those valid in Cartesian coordinates by substituting covariant derivatives for ordinary spatial derivatives and by substituting absolute derivatives for Lagrangian time derivatives.
Note that these replacements have no effect on scalar quantities.
The placement of indices as either superscripts or subscripts is no longer arbitrary --- each term in an equation must transform the same under a change in coordinates.

The absolute derivative of the total strain will appear frequently in the following.
It is trivially obtained from \eqref{eq:depsilonij} by dividing by $dt$
\begin{align}
 \frac{D \e_{ij}}{Dt} &= \frac12 \left(\nabla_i v_j+\nabla_j v_i\right)  \,.
 \label{eq:DepsDt}
\end{align}

Using $d\e_{ij}=d\e^\txt{e}_{ij}+d\e^\txt{p}_{ij}$ and equations \eqref{eq:split-tau}, \eqref{eq:psidot}, \eqref{eq:PR-approx}, and \eqref{eq:DepsDt} we may write the energy conservation equation \eqref{eq:energy-cons} as
\begin{align}
 \rho \pa_t U&=s^{ij}\frac{D\e_{ij}}{Dt}-2\t\pa_t\psi-\bar P\frac{D?\e^i_i?}{Dt} +\rho T\pa_tS \nn\\
 &=\frac12s^{ij}\left(\nabla_i v_j+\nabla_j v_i\right)-2\t\pa_t\psi+\bar P\pa_t\ln\rho +\rho T\pa_tS
 \,,
\end{align}
where we used $d\ln\rho=-d?\e^i_i?$, which follows from mass conservation \eqnref{eq:mass-cons-Lag}; all time derivatives are Lagrangian.

Using equations \eqref{eq:psidot}, \eqref{eq:PR-approx}, \eqref{eq:pure-rot}, \eqref{eq:DepsDt}, \eqref{eq:stress-increments}, and $d\ln\rho=-d?\e^i_i?$, the equations for the mean compressive stress and stress deviators in the Prandtl-Reuss and small anisotropy approximations are
\begin{align}
 \pa_t\bar P&=B\pa_t\ln\rho+\g \rho T\pa_tS\,, \nn\\
  \frac{D s^{ij}}{Dt}
  &= B_s^{ijkl}\frac{D\e_{kl}}{Dt}-B_s^{ijkl}\frac34\frac{s_{kl}}{\t}\pa_t\psi+g^{ij}B\pa_t\ln\rho +\frac12\big(g^{il}s^{jk}-g^{jl}s^{ik}\big)\left(\nabla_k v_l-\nabla_l v_k\right) \nn\\
  &=2G\left(\frac12\left(\nabla^i v^j+\nabla^j v^i\right)+\frac{g^{ij}}{3\rho}\pa_t\rho-\frac{3s^{ij}}{4\t}\pa_t\psi\right)+\frac12\big(g^{il}s^{jk}-g^{jl}s^{ik}\big)\left(\nabla_k v_l-\nabla_l v_k\right)
  \,.
\end{align}

Likewise, equation \eqref{eq:temp-increments} for the temperature becomes
\begin{align}
 \pa_tT&=\g T\pa_t\ln\rho+\frac{T}{C_V}\pa_tS
 \,.
\end{align}

Finally, the entropy production equation is given by \eqref{eq:dS-initial}.
Using $dW^\txt{p}=\t^{ij}d\e^\txt{p}_{ij}=2\t d\psi$ and assuming that the dominant form of heat transport is conduction, i.e. $\rho dQ=-\nabla_{\!i}J^i\approx\kappa\nabla^2Tdt$ (assuming negligible change in the heat flow, $dJ\ll J$ in the relaxation time), we have
\begin{align}
 \rho T \pa_tS&=2\t\pa_t\psi+\kappa\nabla^2T
 \,.
\end{align}
When considering specific geometries we will often neglect the heat transport entirely, and assume $\kappa\nabla^2T=\rho T \pa_tS-2\t\pa_t\psi\approx0$, which will lead to further simplifications.

\section{Summary of the generally invariant equations}

All time derivatives are Lagrangian, i.e. $\pa_t(\ldots)\coleq \pa_t(\ldots)\big|_X$, and the equations below are thus given in mixed Lagrangian/Eulerian form.
Furthermore, we will use the shorthand notation $ (\ldots)_{;i}\coleq\nabla_i(\ldots)$, where $\nabla_i$ denotes the covariant derivative.

{\allowdisplaybreaks
\begin{subequations}\label{eq:geninv}
\begin{align}
\label{eq:geninv-a}
&\textbf{Plastic constitutive:} && \pa_t\psi=\dot\psi(\t,\psi,T,\bar P)
\,, \\
\label{eq:geninv-b}
&\textbf{Mass conservation:} && \pa_t\ln\rho + ?v^i_{;i}?=0 \,, \\
\label{eq:geninv-c}
&\textbf{Momentum conservation:} && \rho \left( \pa_t v^i+\G^i_{jk}v^jv^k \right) =?\t^{ij}_{;j}?
\,,\\
\label{eq:geninv-d}
&\textbf{Stress tensor increments:}\nn\\*
&\qquad\txt{\indent trace part:} && \pa_t\bar P=B\,\pa_t\ln\rho+\g\rho T\pa_tS
\,,\\*
\label{eq:geninv-e}
&\qquad\txt{\indent trace-free part:} && \frac{D?s^i_j?}{Dt}=2G\!\left[\tfrac12\big(?v^i_{;j}?+g^{ik}g_{jl}?v^l_{;k}?\big) +\frac{\d^i_j}{3\rho}\pa_t\rho-\frac{3?s^i_j?}{4\t}\pa_t\psi\right] \nn\\*
&&& \phantom{\frac{D?s^i_j?}{Dt}=}+\frac12\!\left[\big(?v^i_{;n}?-g^{ik}g_{nl}?v^l_{;k}?\big)?s^n_j?-?s^i_n?\big(?v^n_{;j}?-g^{kn}g_{jl}?v^l_{;k}?\big)\right]
\,,\\
\label{eq:geninv-f}
&\textbf{Temperature increment:} && \pa_tT=\g T\pa_t\ln\rho+\frac{T}{C_V}\pa_tS
\,,\\
\label{eq:geninv-g}
&\textbf{Entropy production:} && \rho T\pa_tS=2\t\pa_t\psi+\kappa\nabla^2T \qquad (\text{for } \pa_t J\approx0)
\,,\\
\label{eq:geninv-h}
&\textbf{Energy conservation:} && \rho\pa_t U=\tfrac12\big(?v^i_{;j}?+g^{ik}g_{jl}?v^l_{;k}?\big)?s^j_i? +\bar P\,\pa_t\ln\rho+\rho T\pa_t S-2\t\pa_t\psi
\,,
\end{align}
\end{subequations}
where the absolute derivative in \eqref{eq:geninv-e} is given by
}
\begin{align}
 \frac{D?s^i_j?}{Dt} &= \pa_t ?s^i_j? + \left(\G^i_{lk} ?s^l_j?   - \G^l_{jk} ?s^i_l? \right) v^k   \,,
 \label{eq:dssupisubj}
\end{align}
and $\G^i_{jk}=\tfrac12g^{il}\left(\pa_jg_{lk}+\pa_kg_{lj}-\pa_lg_{jk}\right)$.


In the following sections the generally invariant equations are evaluated in spherical, cylindrical, and spheroidal coordinates.

\section{Spherical symmetry}

In spherical coordinates the line element is $ds^2=dr^2+r^2d\th^2+r^2\sin^2\!\th d\phi^2$ and the non-vanishing Christoffel symbols are:
\begin{align}
 &\G^r_{\th\th}=-r\,, && \G^\th_{\phi\phi}=-\sin\th\cos\th\,, && \G^r_{\phi\phi}=-r\sin\th \,,\nn\\
 &\G^\th_{r\th}=\G^{\phi}_{r\phi}=\frac1r\,, && \G^\phi_{\th\phi}=\cot\th\,.
\end{align}
Spherical symmetry further implies
\begin{align}
 \vv{v}&=(v^r(r),0,0)\,, &
 ?\t^i_j?&=\begin{pmatrix}
            ?\t^r_r?(r) & 0 & 0 \\
            0 & ?\t^\th_\th?(r) & 0 \\
            0 & 0 & ?\t^\th_\th?(r)
           \end{pmatrix},
\end{align}
from which it follows that $?s^\th_\th?=-\frac12?s^r_r?$ and $?\t^\th_\th?=?\t^r_r?-\frac32?s^r_r?$ as well as $\t=\sqrt{\frac{3}{8}?s^i_j??s^j_i?}=\frac34\abs{?s^r_r?}$.

Using these expressions, the spherically symmetric equations follow directly from \eqref{eq:geninv} and thus read
{\allowdisplaybreaks
\begin{subequations}\label{eq:sphere}
\begin{align}
\label{eq:sphere-a}
&\textbf{Plastic constitutive:} && \pa_t\psi=\dot\psi(\t,\psi,T,\bar P)\,,\qquad \bar P=(?s^r_r?-?\t^r_r?)
\,,\\
\label{eq:sphere-b}
&\textbf{Mass conservation:} && \frac1\rho\pa_t\rho + \pa_rv^r+\frac{2v^r}{r}=0
\,,\\
\label{eq:sphere-c}
&\textbf{Momentum conservation:} && \rho \pa_t v^r=\pa_r?\t^r_r?+\frac{3?s^r_r?}{r}
\,,\\
\label{eq:sphere-d}
&\textbf{Stress tensor increments:}
&& \pa_t?\t^r_r?=-\frac{B}{\rho}\pa_t\rho-\frac{3\g\abs{?s^r_r?}}{2}\pa_t\psi+\pa_t?s^r_r?
\,,\\
\label{eq:sphere-e}
& && \pa_t?s^r_r?=\frac{4G}{3}\left(\pa_rv^r-\frac{v^r}{r}-\frac{3}{2}\,\sgn{?s^r_r?}\pa_t\psi\right)
\,,\\
\label{eq:sphere-f}
&\textbf{Temperature increment:} && \pa_tT=\frac{\g T}\rho\pa_t\rho+\frac{3\abs{?s^r_r?}}{2\rho\, C_V}\pa_t\psi
\,,\\
\label{eq:sphere-g}
&\textbf{Energy conservation:} && \rho\pa_t U=-\frac1\rho\pa_t\rho\,?\t^r_r?-\frac{3v^r}{r}?s^r_r?
\,.
\end{align}
\end{subequations}
These equations were derived in~\cite{Preston:1987} for vanishing heat flow $\rho\pa_tQ=\rho T\pa_tS-2\t\pa_t\psi=\kappa\nabla^2T\approx0$, which is justified for situations where solid flow is sufficiently fast that heat transport is negligible compared with plastic work as a source of entropy production.
}

\section{Axisymmetric equations}
The line element for cylindrical coordinates reads $ds^2=dr^2+r^2d\phi^2+dz^2$ and the non-vanishing Christoffel symbols are:
\begin{align}
 \G^r_{\phi\phi}&=-r\,, & \G^\phi_{r\phi}&=\frac1r\,.
\end{align}
Axial symmetry implies
\begin{align}
 \vv{v}&=(v^r(r,z),0,v^z(r,z))\,, &
 ?\t^i_j?&=\begin{pmatrix}
            ?\t^r_r?(r,z) & 0 & ?\t^r_z?(r,z) \\
            0 & ?\t^\ph_\ph?(r,z) & 0 \\
            ?\t^r_z?(r,z) & 0 & ?\t^z_z?(r,z)
           \end{pmatrix}.
\end{align}
Note that since $?\t^r_z?=?s^r_z?$, $\bar P=?s^r_r?-?\t^r_r?$, $?s^\ph_\ph?=-?s^r_r?-?s^z_z?$, and $?\t^\ph_\ph?=?\t^r_r?-2?s^r_r?-?s^z_z?$, we may write all equations in terms of the four variables $?\t^r_r?$, $?s^r_r?$, $?s^z_z?$, $?s^r_z?$.

The {\tpfe} (again for vanishing heat flow where $\rho T\pa_tS\approx2\t\pa_t\psi$) in cylindrical coordinates read:
{\allowdisplaybreaks
\begin{subequations}\label{eq:axi}
\begin{align}
\label{eq:axi-a}
&\textbf{Plastic constitutive:} && \pa_t\psi=\dot\psi(\t,\psi,T,\bar P)\,,\qquad \bar P=(?s^r_r?-?\t^r_r?)
\,,\\
\label{eq:axi-b}
&\textbf{Mass conservation:} && \frac1\rho\pa_t\rho + \pa_r v^r+\frac{v^r}{r}+\pa_zv^z=0
\,,\\
\label{eq:axi-c}
&\textbf{Momentum conservation:} && \rho\pa_tv^r=\pa_r?\t^r_r?+\frac{2?s^r_r?}{r}+\frac{?s^z_z?}{r}+\pa_z?s^r_z? \\
\label{eq:axi-d}
&&& \rho\pa_tv^z=
\pa_r?s^r_z?+\frac1r?s^r_z?+\pa_z\big(?\t^r_r?+?s^z_z?-?s^r_r?\big)
\,,\\
\label{eq:axi-e}
&\textbf{Stress tensor increments:}
&& \pa_t?\t^r_r?=-B\frac1\rho\pa_t\rho-2\t\g\pa_t\psi+\pa_t?s^r_r?
\,,\\
\label{eq:axi-f}
&&& \pa_t?s^r_r?=2G\!\left[\pa_rv^r+\frac{\pa_t\rho}{3\rho}-\frac{3?s^r_r?}{4\t}\pa_t\psi\right]+?s^r_z?\big(\pa_zv^r-\pa_rv^z\big)
\,,\\
\label{eq:axi-g}
&&& \pa_t?s^z_z?=2G\!\left[\pa_zv^z+\frac{\pa_t\rho}{3\rho}-\frac{3?s^z_z?}{4\t}\pa_t\psi\right]+?s^r_z?\big(\pa_rv^z-\pa_zv^r\big)
\,,\\
\label{eq:axi-h}
&&& \pa_t?s^r_z?=G\!\left[\pa_zv^r+\pa_rv^z-\frac{3?s^r_z?}{2\t}\pa_t\psi\right]+\tfrac12\big(?s^z_z?\!-?s^r_r?\big)\big(\pa_zv^r\!-\pa_rv^z\big)
,\\
\label{eq:axi-i}
&\textbf{Temperature increment:} && \pa_tT=\frac{\g T}\rho\pa_t\rho+\frac{2\t}{\rho C_V}\pa_t\psi
\,,\\
\label{eq:axi-j}
&\textbf{Energy conservation:} && \rho\pa_t U
=\pa_zv^z\big(?s^z_z?-?s^r_r?\big)-\frac{v^r}{r}\big(2?s^r_r?+?s^z_z?\big) +?s^r_z?\big(\pa_rv^z+\pa_zv^r\big) \nn\\*
&&& \phantom{\rho\pa_t U=} - ?\t^r_r?\frac1\rho\pa_t\rho
\,,
\end{align}
\end{subequations}
where $2\t=\sqrt{3}\, \sqrt{(?s^r_r?)^2+(?s^z_z?)^2+(?s^r_z?)^2+?s^r_r??s^z_z?}\,$.
To further simplify, we take the sums and differences of \eqref{eq:axi-f}, \eqref{eq:axi-g}, and upon eliminating $\pa_t\rho$ via mass conservation \eqref{eq:axi-b} we find:
}
\begin{align}
 \pa_t?s^r_r?+\pa_t?s^z_z?&=\frac23 G\!\left[\pa_rv^r+\pa_zv^z-\frac{2v^r}{r}-\frac{9}{4\t}\big(?s^r_r?+?s^z_z?\big)\pa_t\psi\right]
 \,,\nn\\
 \pa_t?s^r_r?-\pa_t?s^z_z?&= 2G\!\left[\pa_rv^r-\pa_zv^z+\frac{3}{4\t}\big(?s^r_r?-?s^z_z?\big)\pa_t\psi\right] +2?s^r_z?\big(\pa_zv^r-\pa_rv^z\big)
 \,.
\end{align}

\subsection*{Special case of uniaxial compression:}
In this case the material motion is only in the $z$-direction and the stress tensor has only two independent components~\cite{Wallace:1985}.
Thus
\begin{align}
 \vv{v}&=(0,0,v^z(z))\,, &
 ?\t^i_j?&=\begin{pmatrix}
            ?\t^r_r?(z) & 0 & 0 \\
            0 & ?\t^r_r?(z) & 0 \\
            0 & 0 & ?\t^z_z?(z)
           \end{pmatrix},
\end{align}
where $-?\t^z_z?=\s$ is the normal compressive stress and $?\t^r_r?=2\t-\s$.
It follows that $\bar P=\s-4\t/3$, $?s^z_z?=-4\t/3$, and $?s^r_r?=?s^\ph_\ph?=2\t/3$.
Hence, for uniaxial compression the {\tpfe} reduce to
{\allowdisplaybreaks
\begin{subequations}\label{eq:uniaxi}
\begin{align}
\label{eq:uniaxi-a}
&\textbf{Plastic constitutive:} && \pa_t\psi=\dot\psi(\s,\t,\psi,T)
\,,\\
\label{eq:uniaxi-b}
&\textbf{Mass conservation:} && \frac1\rho\pa_t\rho +\pa_zv^z=0
\,,\\
\label{eq:uniaxi-c}
&\textbf{Momentum conservation:} && \rho\pa_tv^z
=-\pa_z\s
\,,\\
\label{eq:uniaxi-d}
&\textbf{Stress tensor increments:}
&& \pa_t\s=\left[B-\frac43G\right]\!\frac1\rho\pa_t\rho+2\left(G+\t\g\right)\pa_t\psi
\,,\\
\label{eq:uniaxi-e}
&&& \pa_t\t=G\!\left[\frac{\pa_t\rho}{\rho}-\frac{3}{2}\pa_t\psi\right]
\,,\\
\label{eq:uniaxi-f}
&\textbf{Temperature increment:} && \pa_tT=\frac{\g T}\rho\pa_t\rho+\frac{2\t}{\rho C_V}\pa_t\psi
\,,\\
\label{eq:uniaxi-g}
&\textbf{Energy conservation:} && \rho\pa_t U
= \s\frac1\rho\pa_t\rho
\,,
\end{align}
\end{subequations}
in agreement with the equations first obtained by Wallace~\cite{Wallace:1985}.
}

\section{Equations in spheroidal coordinates}
\label{sec:spheroidal}

For problems which deviate from spherical symmetry, spheroidal coordinates may be more useful.
Two cases are of interest: prolate and oblate.
Both lead to very similar expressions for the {\tpfe}.

\subsection{Prolate spheroidal coordinates}
We first consider prolate spheroidal coordinates defined by
\begin{align}
 x&=\sinh\!\m\sin\th\cos\phi\,, \nn\\
 y&=\sinh\!\m\sin\th\sin\phi\,, \nn\\
 z&=\cosh\!\m\cos\th
 \,,
\end{align}
from which $dx$, $dy$, $dz$ in terms of the new coordinates $\m$, $\th$, $\phi$ are easily calculated.
The line element for prolate spheroidal coordinates thus reads
\begin{align}
 ds^2&=h^2(d\m^2+d\th^2)+\sinh^2\!\m\sin^2\!\th d\phi^2 \,,\nn\\
 \textrm{where}\qquad h^2&=\sinh^2\!\m+\sin^2\!\th\nn\\
 &=\cosh^2\!\m-\cos^2\!\th
 \,,
\end{align}
and the non-vanishing Christoffel symbols are:
\begin{align}
 \G^\m_{\m\m}&=\G^\th_{\m\th}=-\G^\m_{\th\th}=\frac{1}{h^2}\cosh\m\sinh\m\,, &
 \G^\th_{\th\th}&=\G^\m_{\m\th}=-\G^\th_{\m\m}=\frac{1}{h^2}\cos\th\sin\th
 \,, \nn\\
 \G^\m_{\phi\phi}&=-\frac{1}{h^2}\sin^2\!\th\sinh\m\cosh\m \,, &
 \G^\th_{\phi\phi}&=-\frac{1}{h^2}\sinh^2\!\m\sin\th\cos\th
 \,, \nn\\
 \G^\phi_{\m\phi}&=\coth\m \,, &
 \G^\phi_{\th\phi}&=\cot\th
 \,.
\end{align}
Axial symmetry implies
\begin{align}
 \vv{v}&=(v^\m(\m,\th),v^\th(\m,\th),0)\,, &
 ?\t^i_j?&=\begin{pmatrix}
            ?\t^\m_\m?(\m,\th) & ?\t^\m_\th?(\m,\th) & 0\\
            ?\t^\m_\th?(\m,\t) & ?\t^\th_\th?(\m,\t) & 0 \\
            0 & 0 & ?\t^\ph_\ph?(\m,\t) 
           \end{pmatrix}.
\end{align}
Since $?\t^\m_\th?=?s^\m_\th?$, $\bar P=?s^\m_\m?-?\t^\m_\m?$, $?s^\ph_\ph?=-?s^\m_\m?-?s^\th_\th?$, and $?\t^\ph_\ph?=?\t^\m_\m?-2?s^\m_\m?-?s^\th_\th?$, we may write all equations in terms of the four variables $?\t^\m_\m?$, $?s^\m_\m?$, $?s^\th_\th?$, $?s^\m_\th?$.
We thus find the {\tpfe} (again for vanishing heat flow where $\rho T\pa_tS\approx2\t\pa_t\psi$):
{\allowdisplaybreaks
\begin{subequations}\label{eq:prol}
\begin{align}
\label{eq:prol-a}
&\textbf{Plastic constitutive:} && \pa_t\psi=\dot\psi(\t,\psi,T,\bar P)\,,\qquad \bar P=?s^\m_\m?-?\t^\m_\m?
\,,\\
\label{eq:prol-b}
&\textbf{Mass conservation:} && -\frac1\rho\pa_t\rho = \pa_\m v^\m+\pa_\th v^\th+\frac{\coth\!\m}{h^2}{(3\sinh^2\!\m+\sin^2\!\th)} v^\m \nn\\
&&& \phantom{-\frac1\rho\pa_t\rho =}\ +\frac{\cot\!\th}{h^2}{(3\sin^2\!\th+\sinh^2\!\m)} v^\th
\,,\\
\label{eq:prol-c}
&\textbf{Momentum conservation:} && \rho\! \left(\!{h^2}\pa_t v^\m+{\sinh\!\m\cosh\!\m\big((v^\m)^2\!-(v^\th)^2\big)+2\sin\th\cos\th v^\m v^\th}\right)=\nn\\*
&&&\hspace*{-5.1cm} =\pa_\m ?\t^\m_\m? +\pa_\th?s^\m_\th?+\coth\!\m\big(2?s^\m_\m?\!+?s^\th_\th?\big)  +\frac{\sinh\!\m\cosh\!\m}{h^2}\big(?s^\m_\m?\!-?s^\th_\th?\big) +\frac{\cot\th}{h^2}\left(3\sin^2\!\th+\sinh^2\!\m\right)?s^\m_\th?
\,,\\
\label{eq:prol-d}
&&& \hspace*{-5.1cm} \rho\! \left(\!{h^2}\pa_t v^\th+{\sin\th\cos\th\big((v^\th)^2\!-(v^\m)^2\big)+2\sinh\!\m\cosh\!\m\, v^\m v^\th}\right)= \nn\\*
&&& \hspace*{-5.1cm} =\pa_\th?\t^\m_\m? +\pa_\m ?s^\m_\th? +\cot\th\big(2?s^\th_\th?+?s^\m_\m?\big) +\Big[\pa_\th+\frac{\sin\th\cos\th}{h^2}\Big]\big(?s^\th_\th?\!-?s^\m_\m?\big) \nn\\*
&&& \hspace*{-5.1cm} \quad +\frac{\coth\!\m}{h^2}\left(3\sinh^2\!\m+\sin^2\!\th\right)?s^\m_\th?
\,,\\
\label{eq:prol-e}
&\textbf{Stress tensor increments:} && \pa_t?\t^\m_\m?=-B\frac1\rho\pa_t\rho-2\t\g\pa_t\psi+\pa_t?s^\m_\m?
\,,\\
\label{eq:prol-f}
& &&\hspace*{-4.2cm} \pa_t?s^\m_\m?
=2G\!\left[\pa_\m v^\m+\frac{\sinh\!\m\cosh\!\m\, v^\m+\sin\!\th\cos\!\th v^\th}{h^2} +\frac{\pa_t\rho}{3\rho}-\frac{3?s^\m_\m?}{4\t}\pa_t\psi\right]
+\big(\pa_\th v^\m-\pa_\m v^\th\big)?s^\m_\th?
\,,\\
\label{eq:prol-g}
& &&\hspace*{-4.2cm} \pa_t?s^\th_\th?
=2G\!\left[\pa_\th v^\th+\frac{\sinh\!\m\cosh\!\m \,v^\m+\sin\!\th\cos\!\th v^\th}{h^2} +\frac{\pa_t\rho}{3\rho}-\frac{3?s^\th_\th?}{4\t}\pa_t\psi\right] 
-\big(\pa_\th v^\m-\pa_\m v^\th\big)?s^\m_\th?
\,,\\
\label{eq:prol-h}
& &&\hspace*{-4.2cm} \pa_t?s^\m_\th?
=2G\!\left[\tfrac12\big(\pa_\m v^\th+\pa_\th v^\m\big) -\frac{3?s^\m_\th?}{4\t}\pa_t\psi\right]
+\frac12\big(?s^\th_\th?-?s^\m_\m?\big)\big(\pa_\th v^\m-\pa_\m v^\th\big)
\,,\\
\label{eq:prol-i}
&\textbf{Temperature increment:} && \pa_tT=\frac{\g T}\rho\pa_t\rho+\frac{2\t}{\rho C_V}\pa_t\psi
\,,\\
\label{eq:prol-j}
&\textbf{Energy conservation:} && \rho\pa_t U=? s^\m_\m?\pa_\m v^\m +?s^\th_\th?\pa_\th v^\th + ? s^\m_\th?\big(\pa_\m v^\th\!+\!\pa_\th v^\m\big) +\big(?s^\m_\m?\!-\!?\t^\m_\m?\big)\frac1\rho\pa_t\rho \nn\\*
&&& \phantom{\rho\pa_t U=} -\frac{\big(?s^\m_\m?\!+ ?s^\th_\th?\big)}{h^2}\Big(\!\coth\!\m\sin^2\!\th\,v^\m+\cot\th\sinh^2\!\m\, v^\th\Big)
\,,
\end{align}
\end{subequations}
where $2\t=\sqrt{3}\, \sqrt{(?s^\m_\m?)^2+(?s^\th_\th?)^2+(?s^\m_\th?)^2+?s^\m_\m??s^\th_\th?}\,$.
Finally, upon using the mass conservation \eqnref{eq:prol-b}, the sums and differences of \eqref{eq:prol-f}, \eqref{eq:prol-g} yield:
\begin{subequations}
\begin{align}
\label{eq:prol-sum}
\pa_t\big(?s^\m_\m?+?s^\th_\th?\big)&=\frac{2G}{3}\!\left[\pa_\m v^\m+\pa_\th v^\th-\frac{2}{h^2}\left(\coth\!\m\sin^2\!\th v^\m+\cot\th\sinh^2\!\m\, v^\th\right)-\frac{9\big(?s^\m_\m?+?s^\th_\th?\big)}{4\t}\pa_t\psi\right]
,\\
\label{eq:prol-diff}
\pa_t\big(?s^\m_\m?-?s^\th_\th?\big)&=2G\left[\pa_\m v^\m-\pa_\th v^\th-\frac{3\big(?s^\m_\m?-?s^\th_\th?\big)}{4\t}\pa_t\psi\right]+2\big(\pa_\th v^\m-\pa_\m v^\th\big)?s^\m_\th?
\,.
\end{align}
\end{subequations}
}

\subsection{Oblate spheroidal coordinates}

Oblate spheroidal coordinates are defined similarly to the previous case
\begin{align}
 x&=\cosh\!\m\cos\th\cos\phi\,, \nn\\
 y&=\cosh\!\m\cos\th\sin\phi\,, \nn\\
 z&=\sinh\!\m\sin\th
 \,,
\end{align}
from which the line element is derived to be
$ds^2=h^2(d\m^2+d\th^2)+\cosh^2\!\m\cos^2\!\th d\phi^2$ with $h^2$ the same as in the previous section.
It differs from the prolate case only by the coefficient of $d\phi^2$.
The non-vanishing Christoffel symbols are:
\begin{align}
 \G^\m_{\m\m}&=\G^\th_{\m\th}=-\G^\m_{\th\th}=\frac1{h^2}{\cosh\m\sinh\m}\,, &
 \G^\th_{\th\th}&=\G^\m_{\m\th}=-\G^\th_{\m\m}=\frac1{h^2}{\cos\th\sin\th}
 \,, \nn\\
 \G^\m_{\phi\phi}&=-\frac1{h^2}{\cos^2\th\sinh\m\cosh\m} \,, &
 \G^\th_{\phi\phi}&=\frac1{h^2}{\cosh^2\m\sin\th\cos\th}
 \,, \nn\\
 \G^\phi_{\m\phi}&=\tanh\m \,, &
 \G^\phi_{\th\phi}&=-\tan\th
 \,.
\end{align}
Note, that the first six (i.e. the entire first line above) are equal to the prolate case; only the last four Christoffel symbols differ.

As expected, the {\tpfe} in oblate spheroidal coordinates are very similar to the prolate case; the only ones that differ are:
{\allowdisplaybreaks
\begin{subequations}\label{eq:obl}
\begin{align}
\label{eq:obl-b}
&\textbf{Mass conservation:} && -\frac1\rho\pa_t\rho = \pa_\m v^\m+\pa_\th v^\th+\frac{\tanh\!\m}{h^2}{(3\cosh^2\!\m-\cos^2\th)} v^\m \nn\\*
&&& \phantom{-\frac1\rho\pa_t\rho =}\ +\frac{\tan\!\th}{h^2}{(3\cos^2\th-\cosh^2\!\m)} v^\th
\,,\\
\label{eq:obl-c}
&\textbf{Momentum conservation:} && \rho\! \left(\!{h^2}\pa_t v^\m+{\sinh\!\m\cosh\!\m\big((v^\m)^2\!-(v^\th)^2\big)+2\sin\th\cos\th v^\m v^\th}\right)=\nn\\*
&&&\hspace*{-5.2cm} =\pa_\m ?\t^\m_\m? +\pa_\th?s^\m_\th?+\tanh\!\m\big(2?s^\m_\m?\!+?s^\th_\th?\big) +\frac{\sinh\!\m\cosh\!\m}{h^2}\big(?s^\m_\m?\!-?s^\th_\th?\big) +\frac{\tan\!\th}{h^2}\left(3\cos^2\!\th-\cosh^2\!\m\right)\!?s^\m_\th?
\,,\\
\label{eq:obl-d}
&&& \hspace*{-5.2cm} \rho\! \left(\!{h^2}\pa_t v^\th+{\sin\th\cos\th\big((v^\th)^2\!-(v^\m)^2\big)+2\sinh\!\m\cosh\!\m\, v^\m v^\th}\right)= \nn\\*
&&& \hspace*{-5.2cm} =\pa_\th?\t^\m_\m?+\pa_\m ?s^\m_\th? -\tan\th\big(2?s^\th_\th?+?s^\m_\m?\big)  +\Big[\pa_\th+\frac{\sin\th\cos\th}{h^2}\Big]\big(?s^\m_\m?\!-?s^\th_\th?\big) \nn\\*
&&& \hspace*{-5.2cm} \quad +\frac{\tanh\!\m}{h^2}\left(3\cosh^2\!\m-\cos^2\!\th\right)?s^\m_\th?
\,,\\
\label{eq:obl-j}
&\textbf{Energy conservation:} && \rho\pa_t U=? s^\m_\m?\pa_\m v^\m +?s^\th_\th?\pa_\th v^\th + ? s^\m_\th?\big(\pa_\m v^\th\!+\!\pa_\th v^\m\big) +\big(?s^\m_\m?\!-\!?\t^\m_\m?\big)\frac1\rho\pa_t\rho \nn\\*
&&& \phantom{\rho\pa_t U=} +\frac{\big(?s^\m_\m?\!+\! ?s^\th_\th?\big)}{h^2}\Big(\!\tanh\!\m\cos^2\!\th\,v^\m+\tan\!\th\cosh^2\!\m\, v^\th\Big)
,
\end{align}
\end{subequations}
whereas the others, i.e. the plastic constitutive, the temperature increment and the stress tensor increments, are equal to the prolate case.
Also \eqref{eq:prol-diff} remains the same, but \eqref{eq:prol-sum} is modified due to the modified mass conservation \eqref{eq:obl-b} and presently reads
}
\begin{align}
\label{eq:obl-sum}
\pa_t\big(?s^\m_\m?+?s^\th_\th?\big)&=\frac{2G}{3}\!\left[\pa_\m v^\m+\pa_\th v^\th+\frac{2}{h^2}\left(\tanh\!\m\cos^2\!\th v^\m+\tan\th\cosh^2\!\m\, v^\th\right)-\frac{9\big(?s^\m_\m?+?s^\th_\th?\big)}{4\t}\pa_t\psi\right]
.
\end{align}

\section{Conclusion}
\label{sec:conclusion}

We have derived the full set of {\tpfe} in general curvilinear coordinates in the Prandtl-Reuss and small anisotropy approximations, summarizing them in eqns.~\eqref{eq:geninv}.
We then presented a number of special cases including spherical symmetry \eqref{eq:sphere}, axisymmetric \eqref{eq:axi}, uniaxial \eqref{eq:uniaxi}, and both prolate and oblate spheroidal geometries in \eqref{eq:prol} and \eqref{eq:obl}.
We chose these geometries for their simplicity as well as for their potential usefulness in applications.
In particular, we envisage that the {\tpfe} in spheroidal coordinates could be used to generalize theories of non-spherical void growth in ductile materials.
Current theories --- see \cite{Gurson:1977,Gologanu:1993,Monchiet:2007,Keralavarma:2017} and references therein for an overview of the literature on this topic --- 
utilize only a subset of the {\tpfe};
changes in temperature, entropy, or stress as functions of time (see eqns.~\eqref{eq:geninv-d}--\eqref{eq:geninv-g} and their spheroidal counter parts in Section~\ref{sec:spheroidal}) have not been taken into account in those theories.
It will be interesting to see the effect of incorporating these additional refinements into models of void growth, a task which we leave to future work.

\subsection*{Acknowledgements}

This work was performed under the auspices of the U.S. Department of Energy under contract DE-AC52-06NA25396.
In particular, the authors are grateful for the support of the Advanced Simulation and Computing, Physics and Engineering Models Program.

\bibliographystyle{utphys-custom}
\bibliography{dislocations}

\end{document}